\documentclass[aip,superscriptaddress,preprint,apl]{revtex4}
\usepackage{amsmath,amsfonts,amssymb,natbib}
\usepackage[english]{babel}
\usepackage{booktabs,longtable}
\usepackage{graphicx}
\usepackage{dcolumn}
\usepackage{bm}
\usepackage[T1]{fontenc}
\usepackage[latin9]{inputenc}
\usepackage{color}
\usepackage{textcomp}
\usepackage{amsmath}

\begin{document}

\title{Van der Waals solid phase epitaxy to grow large-area manganese-doped MoSe$_{2}$ few-layers on SiO$_{2}$/Si}

\author{C. Vergnaud}
\affiliation{Univ. Grenoble Alpes, CEA, CNRS, Grenoble INP, IRIG-SPINTEC, 38000 Grenoble, France}

\author{M. Gay}
\affiliation{Univ. Grenoble Alpes, CEA, LETI, Minatec Campus, 38000 Grenoble, France}

\author{C. Alvarez}
\affiliation{Univ. Grenoble Alpes, CEA, IRIG-MEM, 38000 Grenoble, France}

\author{M.-T. Dau}
\affiliation{Univ. Grenoble Alpes, CEA, CNRS, Grenoble INP, IRIG-SPINTEC, 38000 Grenoble, France}

\author{F. Pierre}
\affiliation{Univ. Grenoble Alpes, CEA, LETI, Minatec Campus, 38000 Grenoble, France}

\author{D. Jalabert}
\affiliation{Univ. Grenoble Alpes, CEA, IRIG-MEM, 38000 Grenoble, France}

\author{C. Licitra}
\affiliation{Univ. Grenoble Alpes, CEA, LETI, Minatec Campus, 38000 Grenoble, France}

\author{A. Marty}
\affiliation{Univ. Grenoble Alpes, CEA, CNRS, Grenoble INP, IRIG-SPINTEC, 38000 Grenoble, France}

\author{C. Beign\'e}
\affiliation{Univ. Grenoble Alpes, CEA, CNRS, Grenoble INP, IRIG-SPINTEC, 38000 Grenoble, France}

\author{B. Gr\'evin}
\affiliation{Univ. Grenoble Alpes, CEA, IRIG-SYMMES, 38000 Grenoble, France}

\author{O. Renault}
\affiliation{Univ. Grenoble Alpes, CEA, LETI, Minatec Campus, 38000 Grenoble, France}

\author{H. Okuno}
\affiliation{Univ. Grenoble Alpes, CEA, IRIG-MEM, 38000 Grenoble, France}

\author{M. Jamet}
\email{matthieu.jamet@cea.fr}
\affiliation{Univ. Grenoble Alpes, CEA, CNRS, Grenoble INP, IRIG-SPINTEC, 38000 Grenoble, France}

\date{\today}

\begin{abstract}
Large-area growth of continuous transition metal dichalcogenides (TMDCs) layers is a prerequisite to transfer their exceptional electronic and optical properties into practical devices. It still represents an open issue nowadays. Electric and magnetic doping of TMDC layers to develop basic devices such as p-n junctions or diluted magnetic semiconductors for spintronic applications are also an important field of investigation. Here, we have developed two different techniques to grow MoSe$_2$ mono- and multi-layers on SiO$_2$/Si substrates over large areas. First, we co-deposited Mo and Se atoms on SiO$_2$/Si by molecular beam epitaxy in the van der Waals regime to obtain continuous MoSe$_2$ monolayers over 1 cm$^2$. To grow MoSe$_2$ multilayers, we then used the van der Waals solid phase epitaxy which consists in depositing an amorphous Se/Mo bilayer on top of a co-deposited MoSe$_2$ monolayer which serves as a van der Waals growth template. By annealing, we obtained continuous MoSe$_2$ multilayers over 1 cm$^2$. Moreover, by inserting a thin layer of Mn in the stack, we could demonstrate the incorporation of up to 10 \% of Mn in MoSe$_2$ bilayers.
\end{abstract}

\maketitle

%
%
%
%

\section{Introduction}

Recently, there has been tremendous interest in layered transition metal dichalcogenides (TMDCs). It started with the discovery that the canonical TMDC MoS$_2$ becomes a direct bandgap semiconductor when isolated in monolayer (ML) form \cite{Splendiani2010,Mak2010}. Many of the TMDCs have since been shown to exhibit a strong excitonic feature, showing strong and spectrally narrow photoluminescence and absorption making them very good candidates for novel optoelectronic devices \cite{Lopez-Sanchez2013,Gutierrez2013}. The combination of a bandgap (1-2 eV) and an atomically thin conducting channel also makes TMDCs ideal materials for next generation field effect transitors in microelectronics \cite{Radisavljevic2011}. Finally, in the monolayer form, due to the breaking of inversion symmetry and strong spin-orbit coupling, the carriers in TMDCs are with a valley index and this intrinsic degree of freedom of electrons can be exploited to develop novel spintronic and valleytronic devices \cite{Xiao2012,Zeng2012,Mak2012}. Mechanical exfoliation \cite{Geim2007} is one of the most popular methods to produce monolayer flakes, and though it is possible to produce very high quality flakes in this manner, the samples remain very small ($<$10 $\mu$m for TMDCs). So far, the highest quality flakes are produced by mechanical exfoliation and subsequent encapsulation in hBN \cite{Cadiz2017,Wierzbowski2017}. Typically, encapsulation leads to even smaller flake sizes and involves very long process. Monolayers can also be directly grown on various substrates, typically by chemical vapor deposition (CVD) or metal organic CVD (MOCVD) (see reviews \cite{Cai2018,Li2018,Qi2018,Song2017,Choi2017a}). These techniques lead to much larger flakes, up to 100 $\mu$m single crystal in the best cases and continuous wafer-scale polycrystalline by MOCVD \cite{Kang2015}. However, the geological TMDC crystals that are exfoliated, along with those produced by CVD and MOCVD, have been shown to have a high density of structural defects, high impurity levels, and a large degree of variability across the same sample surface. In this respect, molecular beam epitaxy (MBE) is a promising growth method for TMDCs, and heterostructures thereof, due to the potential of enhanced quality provided by a combination of high purity elemental sources and growth in an ultrahigh vacuum (UHV) system. Moreover, MBE offers more flexibility in the choice of the substrate and the transition metal \cite{Zhang2013,Ugeda2014,Dau2017,Dau2018,Vishwanath2015,Lehtinen2015,Liu2015,Jiao2015,Roy2016,Chen2017,Yue2017,Wei2017,Choi2017}. It also allows to stack different TMDCs and grow complex heterostructures with high crystalline quality \cite{Xenogiannopoulou2015}. The growth temperature by MBE being lower than that for CVD, it favors nucleation and leads to high layer coverage close to 100 \% and a good control of the number of deposited layers. The MBE growth of TMDCs is usually performed in the van der Waals regime \cite{Koma1992}. In this regime, the interaction between the substrate and the epilayer is weak enough that it releases the constraints of lattice matching \cite{Koma1999}. In this Letter, we report the growth of mono and multilayers of MoSe$_2$ on SiO$_2$(300 nm)/Si substrates on large areas (cm$^{2}$). We show that proper surface treatment and growth conditions allow for the growth of one monolayer of MoSe$_2$ with high coverage ($>$93 \%) by co-depositing Mo and Se atoms. However, when approaching the 100 \% coverage and when the second layer starts to grow, we observe the starting of vertical three-dimensional growth at some grain boundaries preventing the growth of a perfecly flat two-dimensional MoSe$_2$ film thicker than one monolayer. To circumvent this issue, we developed an original method based on the solid phase epitaxy (SPE) concept. It consists in depositing the Se layer and the Mo layer on top at low temperature followed by annealing to form MoSe$_2$. However, we could show that two-dimensional MoSe$_2$ multilayers only form when SPE is conducted on a MoSe$_2$ template monolayer grown by co-deposition. We then call this technique van der Waals solid phase epitaxy (vdW-SPE). By this, we could grow up to 3 ML-thick MoSe$_2$ films as well as Mn-doped MoSe$_2$ multilayers. Despite the small size of individual grains (10 to 50 nm) that may affect the carrier mobility in electronic devices, both the co-deposited MoSe$_2$ monolayer and the vdW-SPE grown multilayers exhibit good quality on SiO$_2$ over large areas with structural, electronic and optical properties comparable to those of MoSe$_2$ exfoliated flakes.

\section{\label{codep} Growth of MoSe$_2$ by co-deposition}

\begin{figure}[hbtp]
\centering
\includegraphics[scale=0.5]{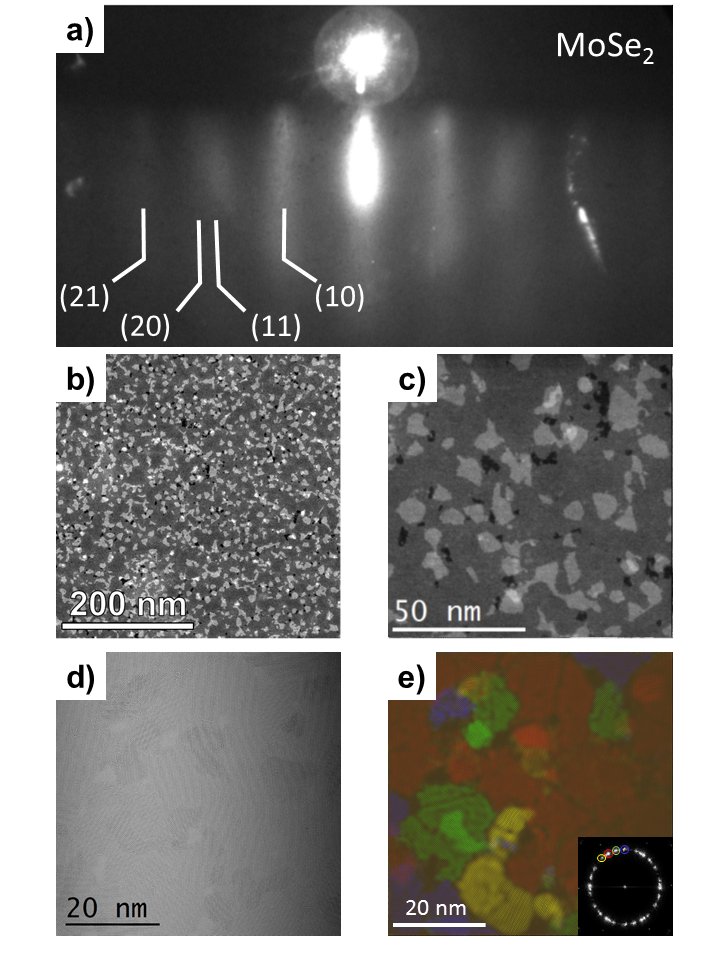}
\caption{\textit{(a)}, RHEED pattern of 1.2 ML of MoSe$_2$ deposited on SiO$_2$(300 nm)/Si. \textit{(b,c)}, STEM images at two different magnifications. We observe 3 different contrasts: black is for bare graphene, dark grey is for 1 ML of MoSe$_2$ and light grey for 2 MLs of MoSe$_2$. \textit{(d)}, High resolution TEM image of 1.2 ML of MoSe$_2$. \textit{(e)}, Representation of individual MoSe$_2$ domains using false colors. Inset: Fourier transform of the TEM image in \textit{(d)}. Each color corresponds to a domain orientation.}\label{Fig1}
\end{figure}

Here, we describe the growth of one monolayer of MoSe$_2$ with high surface coverage on SiO$_2$. The film growth is performed in ultrahigh vacuum (base pressure in the low 10$^{-10}$ mbar) by evaporating molybdenum using an electron gun evaporator and selenium using a standard Knudsen cell. The SiO$_2$(300 nm)/Si substrate is first dipped during 5 minutes into a piranha solution consisting of 3 parts of concentrated sulfuric acid and 1 part of hydrogen peroxide. This mixture removes all the organic residues off the substrate and leads to a hydroxylic surface termination by adding OH groups \cite{Helms1993}. The substrate is then rinsed in deionized water, dried with clean nitrogen gas and immediately introduced into the MBE machine. Its temperature is rised slowly during 45 minutes up to the deposition temperature of 600$^{\circ}$C. In order to limit the formation of Se vacancies, we saturate the surface with Se by using a high Se:Mo ratio of 15:1. The Se partial pressure is about 2$\times$10$^{-6}$ mbar as measured by a retractable Bayard-Alpert gauge in place of the sample. The MoSe$_2$ deposition rate is then fully determined by the Mo evaporation rate monitored by a quartz balance and fixed at 0.3 \AA/min. After the deposition of 1.2 ML, the film is annealed at 750$^{\circ}$C during 15 minutes under Se flux.
This annealing procedure is necessary to improve the crystalline quality within the MoSe$_2$ grains. However due to the decomposition of MoSe$_2$ at high temperatures (Se atoms starting to desorb), we have to use a counter-flux of selenium to limit this desorption and MoSe$_2$ decomposition. An example of crystalline quality improvement is given in Ref. \cite{Alvarez2018} where we could clearly see that mirror-twin boundaries (commonly observed in the vdW epitaxy of MoSe$_2$) can be almost suppressed by this annealing procedure. However, since mirror-twin boundaries reflect a deficit of selenium, Mo-rich areas start to form within the film when annealing the layer without Se flux. The Se counter-flux is thus necessary to keep the right stoichiometry and avoid the formation of those Mo-rich regions. The reflection high energy electron diffraction (RHEED) pattern is initially diffuse due to the amorphous character of SiO$_2$ and progressively evolves into a streaky pattern during the growth. The final RHEED pattern is shown in Fig.~\ref{Fig1}a and remains unchanged when changing the azimutal angle. This isotropic character is indicative of an in-plane polycrystalline structure. For such a two-dimensional film with in-plane polycrystalline character, the reciprocal space consists in concentric cylinders and its intersection with the Ewald sphere gives isotropic streaks as found experimentally. 

\begin{figure}[hbtp]
\centering
\includegraphics[scale=0.2]{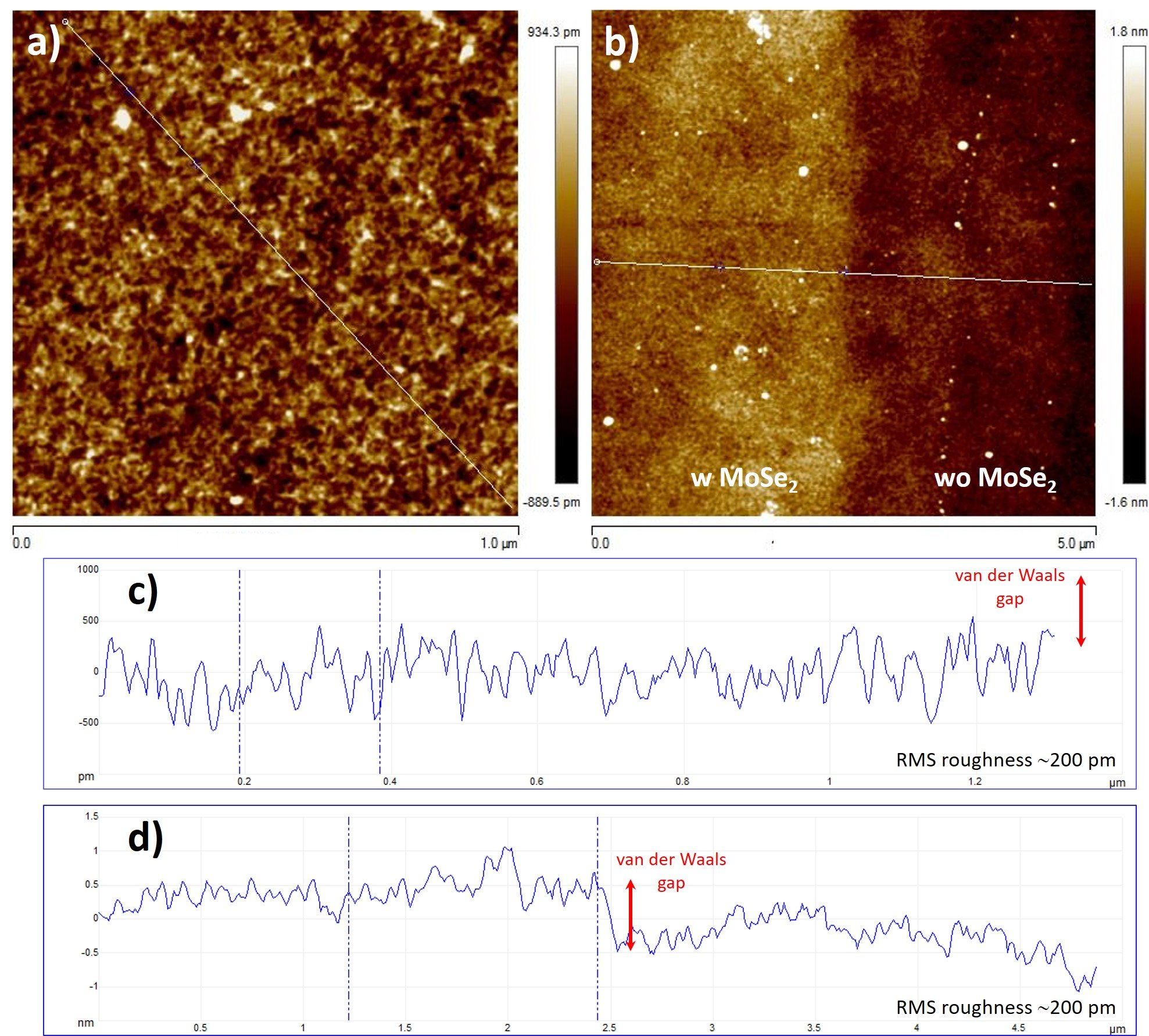}
\caption{\textit{(a)} 1$\times$1 $\mu$m$^{2}$ and \textit{(b)} 5$\times$5 $\mu$m$^{2}$ AFM images of 1 ML of MoSe$_2$ grown by co-deposition on SiO$_2$/Si. \textit{(c)} and \textit{(d)} Height profiles recorded along the white solid lines in \textit{(a)} and \textit{(b)} respectively. The root mean square (RMS) roughness and the expected van der Waals gap (650 pm) between MoSe$_2$ and SiO$_2$ are indicated in the height profiles.}\label{Fig2}
\end{figure}
\begin{figure}[hbtp]
\centering
\includegraphics[scale=0.5]{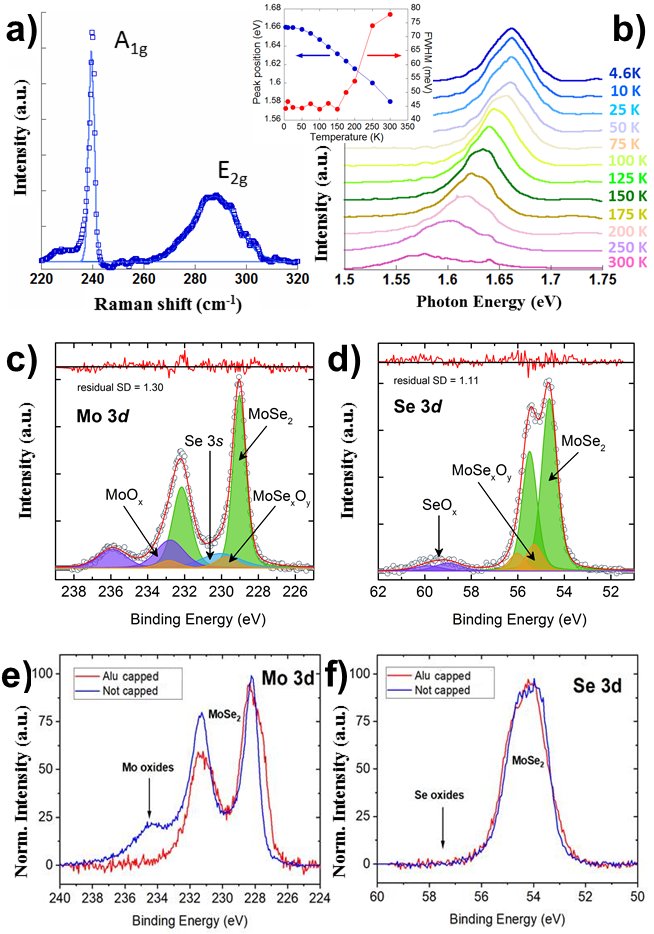}
\caption{\textit{(a)}, Typical Raman spectrum obtained on 1.2 ML of MoSe$_2$. The blue solid line corresponds to a fit of the $A_{1g}$ peak. \textit{(b)}, Photoluminescence signal as a function of temperature. Inset: position and FWHM of PL peaks as a function of temperature. XPS Mo 3d \textit{(c)} and Se 3d \textit{(d)} high-resolution core level spectra. (e) and (f) XPS core level spectra of Mo 3d and Se 3d respectively for 1.2 ML of MoSe$_2$ with and without aluminum capping.}\label{Fig3}
\end{figure}

STEM images of the film are shown in Fig.~\ref{Fig1}b and \ref{Fig1}c where the contrast correlates with the number of MoSe$_2$ layers. Black areas are for 0 ML of MoSe$_2$ and correspond to the graphene underlayer as given by complementary high-resolution TEM images. Then the contrast becomes brighter as the number of MoSe$_2$ layers increases. Dark grey areas are for 1 ML of MoSe$_2$ and we find a surface coverage $>$ 93 \%. Finally, light grey areas are for 2 ML of MoSe$_2$ and they represent $\approx$17 \% of coverage. The total thickness of MoSe$_2$ estimated from STEM images is $\approx$1.2 ML which corresponds to the thickness estimated by the quartz balance. A TEM image is shown in Fig.~\ref{Fig1}d, we can clearly see the atomic arrangements. In Fig.~\ref{Fig1}e, since individual grains are randomly oriented in-plane, they can be selected in the Fourier transform of the TEM image (see the inset) and visualized in false color. From this analysis, we deduce the mean grain size between 10 and 50 nm. Atomic force microscopy (AFM) in Fig.~\ref{Fig2}a and ~\ref{Fig2}b confirms the uniformity of the MoSe$_2$ layer over micrometers. We find a root-mean-square (RMS) roughness of the order of 200 pm which corresponds to the one of the pristine SiO$_2$/Si substrate. By selecting an uncovered area in Fig.~\ref{Fig2}b, we deduce in the height profile of Fig.~\ref{Fig2}d the distance between the MoSe$_2$ layer and the SiO$_2$ surface of the order of 650 pm which corresponds to the van der Waals gap of MoSe$_2$. A typical Raman spectrum is shown in Fig.~\ref{Fig3}a after subtraction of the substrate signal. We clearly see two main peaks corresponding to the $A_{1g}$ at 239.8 cm$^{-1}$ and $E_{2g}$ at 287.4 cm$^{-1}$ vibration modes of MoSe$_2$. The $A_{1g}$ peak corresponds to out-of-plane vibration mode and the $E_{2g}$ to in-plane vibration mode.


The peaks are at the same positions as those of CVD-grown and exfoliated MoSe$_2$ monolayer \cite{Tonndorf2013,Zhao2017,Xia2014} confirming the good quality of the MBE-grown MoSe$_2$ monolayer. This is supported by the $A_{1g}$ peak width of the order of 2.5 cm$^{-1}$ which is comparable to the one of exfoliated flakes ($\approx$2 cm$^{-1}$) \cite{Tonndorf2013}. We also performed Raman maps over 20$\times$20 $\mu$m$^{2}$ areas with a 2 $\mu$m step, the spot size being of the order of 1 $\mu$m. All the recorded peaks fall at the same position and exhibit the same intensity within 5 \% showing the good uniformity of the MoSe$_2$ layer over large areas. Photoluminescence (PL) spectra are also shown in Fig.~\ref{Fig3}b as a function of temperature. When decreasing the temperature, the PL signal increases and the peak is blueshifted from 1.58 eV at room temperature to 1.66 eV at 4.6 K. The peak position and full width at half maximum (FWHM) as a function of temperature are reported in the inset of Fig.~\ref{Fig3}b. The peak positions are in quantitative agreement with previous PL measurements on MoSe$_2$ flakes \cite{Ross2013,Tonndorf2013} and correspond to the emission of the neutral exciton $X^{0}$. Concerning the peak FWHM, it varies from $\approx$40 meV at low temperature up to $\approx$80 meV at room temperature. At room temperature, Tonndorf \textit{et al.} also found a single emission peak at the same energy position with a width of $\approx$60 meV close to our value. Surprisingly, the FWHM remains constant below 150 K while it should still decrease down to 4.6 K. This observation is in agreement with the presence of the negatively charged exciton $X^{-}$ (trion) emission at low temperature as shown in Ref. \cite{Ross2013}. The presence of this additional peak leads to a broadening of the neutral exciton peak at low temperature by adding a shoulder at lower energy (see for instance the PL spectrum at 50 K in Fig.~\ref{Fig3}b). We cannot resolve individually the $X^{0}$ and $X^{-}$ peaks due to their large width which is probably a consequence of short-scale fluctuations and oxidation (see photoemission measurements below) in the MoSe$_2$ layer. To fit the emission peak at 4.6 K with two individual peaks (for the negatively charged and neutral excitons), we need to consider a linewidth of $\approx$30 meV for each. For comparison, Wang \textit{et al.} found a linewidth of $\approx$20 meV in bulk-exfoliated flakes which is comparable to our value \cite{Wang2015}. At higher temperature, the negatively charged exciton peak disappears because electrons can escape their bound trion state due to thermal fluctuations. The presence of the negative trion emission can be due to electrostatic doping by the SiO$_2$ substrate (charges being photocreated at the surface of SiO$_2$) or to intrinsic defects such as Se vacancies inducing n-type doping. We further investigated the electronic state of Mo and Se by XPS. Fig.~\ref{Fig3}c (resp. Fig.~\ref{Fig3}d) presents the Mo 3d (resp. Se 3d) high-resolution spectra from 1.2 ML of MoSe$_2$. Each spectral fitting reveals a certain number of doublets. The relative weights of each doublet are given in Table~\ref{Table1}.

\begin{table}[htt!!!]
\caption{\label{Table1}Results of XPS spectra fits for the three samples: 1.2 ML MoSe$_2$ by co-deposition; 3 ML MoSe$_2$ by vdW-SPE and 3 ML MoSe$_2$ by vdW-SPE doped with 10 \% Mn. The binding energy (BE) and full width at half maximum (FWHM) are reported for each core level (CL) in eV. The contribution of each phase (MoSe$_2$, MoSe$_x$O$_y$, MoO$_x$ and SeO$_x$) are given in \%.}
\begin{ruledtabular}
\begin{tabular}{lclclclclclclclclclclcl}
   \multicolumn{2}{c}{} & \multicolumn{3}{c}{MoSe$_2$} & \multicolumn{3}{c}{MoSe$_x$O$_y$} & \multicolumn{3}{c}{MoO$_x$ \& SeO$_x$} \\
   Sample & CL & BE & FWHM & \% & BE & FWHM & \% & BE & FWHM & \% \\
   \hline
   1.2 ML & Mo 3d$_{5/2}$ & 229.0 & 0.8 & 72 & 229.7 & 1.4 & 3 & 232.8 & 1.8 & 25 \\
   co-dep. & Se 3d$_{5/2}$ & 54.6 & 0.8 & 80 & 55.3 & 0.8 & 13 & 59.0 & 1.4 & 7 \\
   \hline
   3 ML & Mo 3d$_{5/2}$ & 228.9 & 0.7 & 78 & 229.6 & 1.0 & 3 & 232.9 & 1.8 & 19 \\
   SPE & Se 3d$_{5/2}$ & 54.5 & 0.7 & 84 & 55.1 & 0.7 & 10 & 59.0 & 1.5 & 6 \\
   \hline
   3 ML & Mo 3d$_{5/2}$ & 228.9 & 0.7 & 79 & 229.6 & 1.2 & 5 & 232.5 & 2.3 & 16 \\
   SPE & Se 3d$_{5/2}$ & 54.5 & 0.7 & 81 & 55.1 & 0.7 & 16 & 58.9 & 1.3 & 3 \\
   10\% Mn & Mn 2p$_{3/2}$ & 641.2 & 3.0 & - & - & - & - & - & - & - \\
\end{tabular}
\end{ruledtabular}
\end{table}

The main doublet in green is assigned to stoichiometric MoSe$_2$, i.e. Mo$^{4+}$ and Se$^{2-}$ oxidation states. Mo 3d$_{5/2}$ binding energy is 229.0 eV. The corresponding Se 3d$_{5/2}$ binding energy is 54.6 eV. These values are in good agreement with the MoSe$_2$ binding energies reported in the literature \cite{Vishwanath2015}. Concerning the Mo 3d lines, we have to take into account the Se 3s peak at 230.1 eV to fit correctly the whole spectrum. As revealed by TEM measurements, the 1.2 ML MoSe$_2$ film is not perfectly homogeneous and thus locally presents a different chemical environment for Mo and Se atoms. As a result, vacancies, grain boundaries and edge-related defects may have distinct fingerprints in the XPS spectra as secondary doublets next to the main one, related to stoichiometric MoSe$_2$. In addition, as the sample has been exposed to air prior to the analysis, the defects have been partly oxidized and appear at higher binding energies. The orange doublet is attributed to oxidized defects and is accordingly referred to as MoSe$_x$O$_y$ with associated binding energies for Mo 3d and Se 3d of 229.7 eV and 55.3 eV respectively. These doublets present a 0.7 eV shift to higher binding energies for Mo and Se 3d. They represent 3 \% and 13 \% of the total amount of Mo and Se respectively. We attribute this MoSe$_x$O$_y$ component to partial oxidation within the grains at some point defects such as Se vacancies. The purple doublet is assigned to Mo and Se oxides MoO$_x$ and SeO$_x$ at 232.8 eV and 59.0 eV respectively. Indeed these binding energies are close to the one of Mo 3d$_{5/2}$ in MoO$_3$ and Se 3d$_{5/2}$ in SeO$_2$ respectively \cite{Choi1996,Spevack1992,Baltrusaitis2015}. The presence of such a doublet is due to ageing in air. This is confirmed by XPS measurements on the same 1.2 ML of MoSe$_2$ capped with 2 nm of aluminum showing no trace of oxidation in Fig.~\ref{Fig3}e and ~\ref{Fig3}f. MoO$_x$ and SeO$_x$ components can be attributed to oxidized Mo and Se at the edges of grains and at highly disordered grain boundaries. Comparing the percentages of MoO$_x$ (25 \%) and SeO$_x$ (7 \%), we can conclude that grain edges and disordered boundaries are Mo rich. 

\begin{figure}[hbtp]
\centering
\includegraphics[scale=0.4]{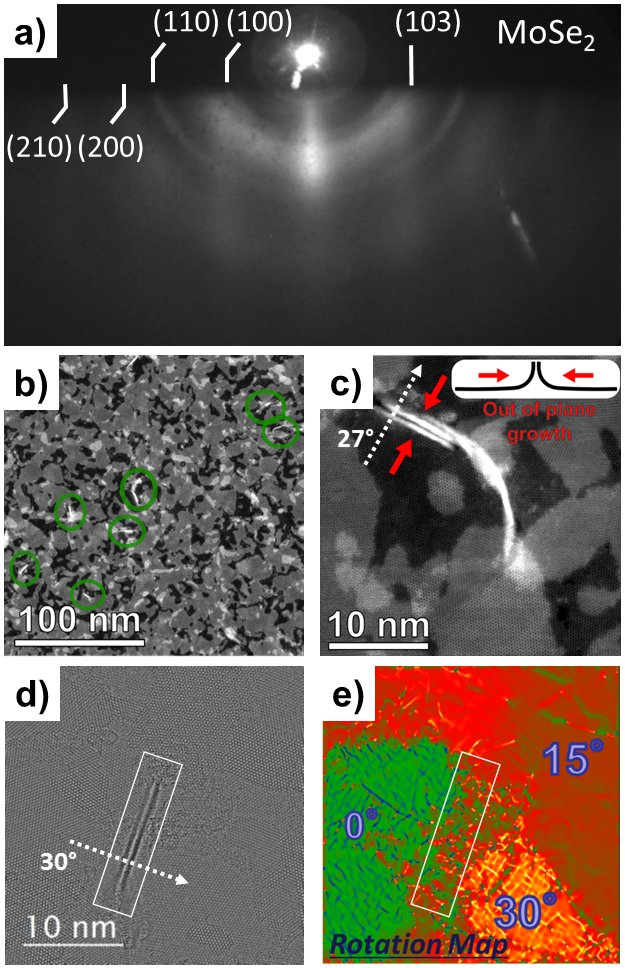}
\caption{\textit{(a)}, RHEED pattern of $\approx$2 ML of MoSe$_2$ grown by co-depositing Mo and Se. \textit{(b)}, STEM image of $\approx$2 ML of MoSe$_2$. Vertical growths appear as white lines in the image and are highlighted by green circles. Black areas correspond to the first MoSe$_2$ layer, dark grey areas to the second layer and the light grey to the third layer starting to grow. \textit{(c)}, Zoom-in of the STEM image showing the three-dimensional growth of the first MoSe$_2$ layer at the junction of two grains. Inset: sketch of the vertical growth process. The relative angle between the two nearby grains is $\approx$27$^{\circ}$. \textit{(d)}, High resolution TEM image showing vertical growth of the first MoSe$_2$ layer highlighted in the white box. The relative angle between the two nearby grains is 30$^{\circ}$ as shown by the associated rotation map in \textit{(e)} where the color scale gives an absolute value of the grains orientation.}\label{Fig4}
\end{figure}

When increasing the film thickness up to $\approx$2 ML in the same growth conditions, we systematically observe the formation of rings in the RHEED pattern in Fig.~\ref{Fig4}a. Those rings come from the three-dimensional growth of MoSe$_2$. In Fig.~\ref{Fig4}b and ~\ref{Fig4}c, STEM images reveal the starting of vertical growth at the junction between two grains as illustrated in the inset of Fig.~\ref{Fig4}c. Due to their random orientation, additional spheres appear in the reciprocal space giving rise to rings in the RHEED pattern as they intersect the Ewald sphere. In order to observe the out-of-plane growth, the total coverage needs to be close to 2 ML (we do not observe it for 1.2 ML) whereas it mainly takes place within the first MoSe$_2$ layer. Vishwanath \textit{et al.} already observed this out-of-plane growth in the case of MoSe$_2$ multilayers grown on HOPG \cite{Vishwanath2015}. They call them protrusions and interpreted their formation as due to the interaction between two MoSe$_2$ domains growing on both sides of HOPG high steps. They ruled out the effect of a difference of thermal expansion (the growth being performed at high temperature) between MoSe$_2$ and the HOPG substrate. We can conclude the same in our case since the thermal expansion of SiO$_2$ (0.56$\times$10$^{-6}$ $^{\circ}$C$^{-1}$ \cite{Elkareh1995}) is negligible as compared to that of MoSe$_2$ (7.24$\times$10$^{-6}$ $^{\circ}$C$^{-1}$ \cite{Elmahalawy1976}). Instead of out-of-plane growth, we should observe the formation of cracks in the MoSe$_2$ layer when the system is cooling down. Since MoSe$_2$ grains are randomly oriented in-plane, we rather believe that, for some specific relative orientations of the grains, the system cannot accomodate a stable grain boundary in which defects would cost too much energy. This effect may also be strengthened by the low mobility of Mo and Se atoms at the surface of SiO$_2$ as compared to the mobility on graphene \cite{Alvarez2018}. At those grain junctions, the system is driven to vertical growth and this process is even accelerated because it is kinetically favorable \cite{Kong2013}. By analysing high resolution TEM images (see Fig.~\ref{Fig4}d and ~\ref{Fig4}e), we find that the relative grain orientations leading to vertical growth are mostly falling in the 25$^{\circ}$-35$^{\circ}$ range. It indeed corresponds to the most unfavorable relative crystal orientation to form a grain boundary. However, we could find other orientations showing that this vertical growth may be the consequence of several mechanisms. In particular, the surface roughness of SiO$_2$ and the presence of residual molecules and defects may also favor out-of-plane and vertical growth \cite{Zheng2017}. We stress the fact that we always obtained the same results when varying the growth rate and temperature showing that vertical growth is almost independent on the growth parameters. In order to grow two-dimensional MoSe$_2$ multilayers, we developed an original method based on the solid phase epitaxy concept described in the following.

\section{Growth of MoSe$_2$ by van der Waals-solid phase epitaxy}\label{SPE}

Solid-phase epitaxy (SPE) corresponds to a transition between the amorphous and crystalline phases of a material. It is usually done by first depositing a film of amorphous material on a crystalline substrate. The substrate is then heated to crystallize the film. The single crystal substrate serves here as a template for crystal growth \cite{SPE}. 

\begin{figure}[hbtp]
\centering
\includegraphics[scale=0.4]{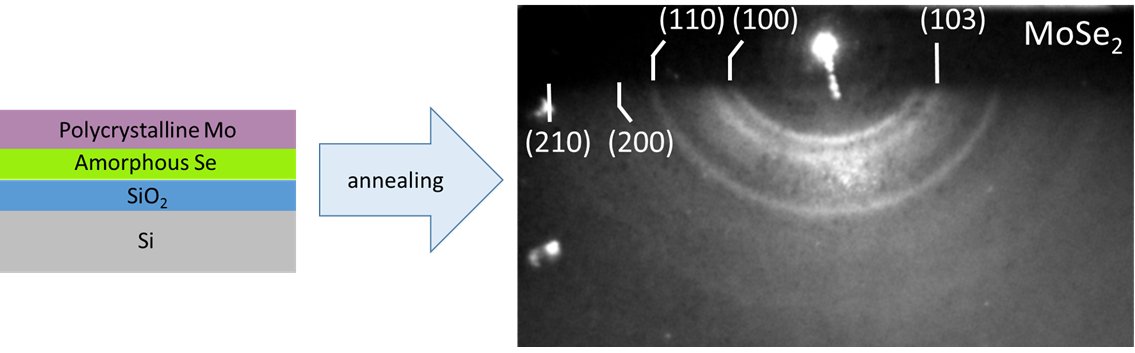}
\caption{Sketch of the initial stack for SPE growth on SiO$_2$ and corresponding final RHEED pattern after annealing. The RHEED pattern is initially diffuse showing the amorphous character of the stack.}\label{Fig5}
\end{figure}
\begin{figure}[hbtp]
\centering
\includegraphics[scale=0.5]{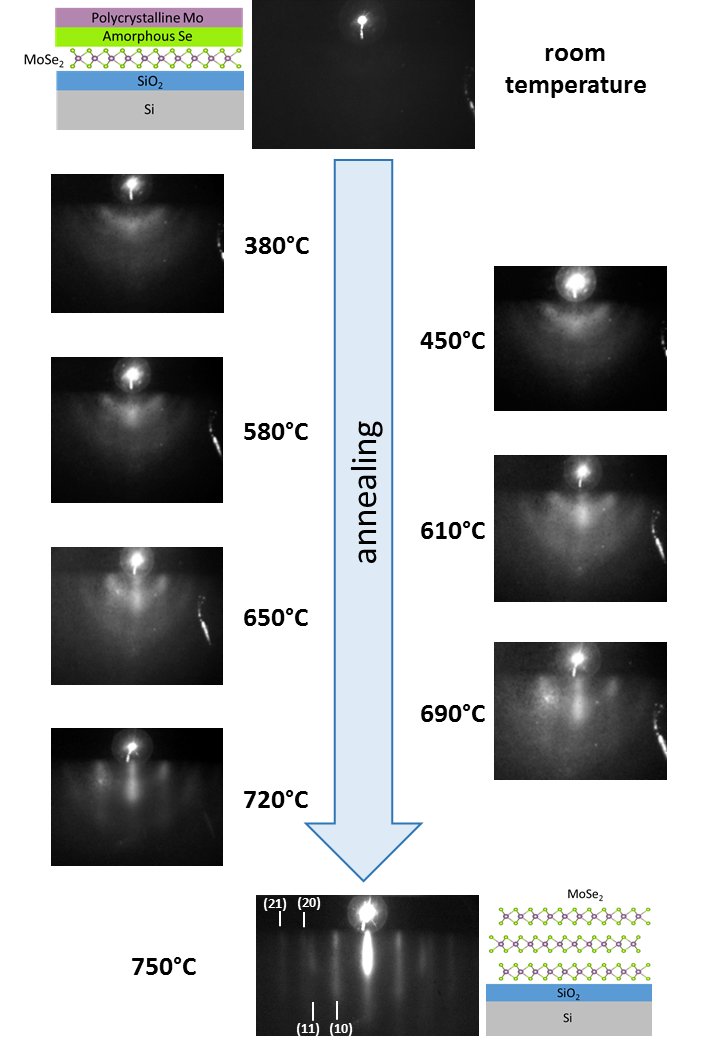}
\caption{Evolution of the RHEED pattern during annealing for the SPE growth on a MoSe$_2$ co-deposited monolayer.}\label{Fig6}
\end{figure}
\begin{figure}[hbtp]
\centering
\includegraphics[scale=0.6]{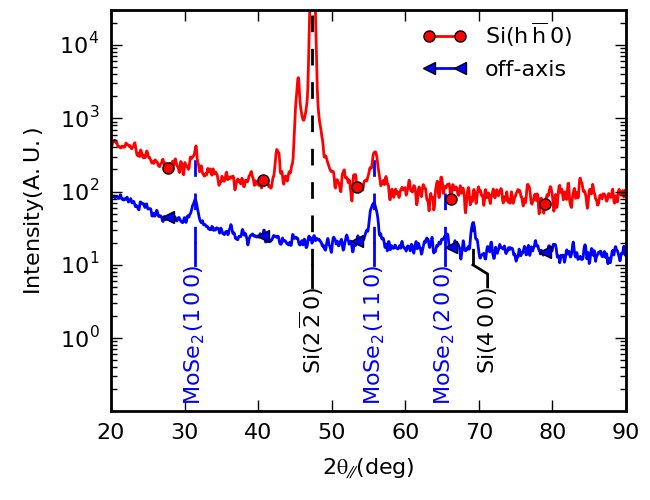}
\caption{Grazing x-ray diffraction spectra of 2 ML of MoSe$_2$ grown by vdW-SPE on 1 ML MoSe$_2$ grown by co-deposition along two different radial reciprocal space directions showing the isotropic character of the film. The vertical dotted lines indicate the different peak positions.}\label{Fig7}
\end{figure}

In our case, since the SiO$_2$ substrate is amorphous, we expect to obtain two-dimensional polycrystalline MoSe$_2$ multilayers. For this purpose, we first deposit a $\approx$10 nm-thick amorphous selenium film and 0.6 nm of Mo on top at room temperature in order to keep a high Se:Mo ratio of $\approx$15. 0.6 nm of Mo correspond to 2 ML of MoSe$_2$ when selenized. The RHEED pattern is initially diffuse. We then annealed this stack and rings start to appear in the RHEED pattern at 600$^{\circ}$C as shown in Fig.~\ref{Fig5}. Above this temperature, the RHEED pattern does not evolve anymore up to 800$^{\circ}$C. We conclude that MoSe$_2$ grains are growing with random orientations, the growth is three-dimensional. In order to favor two-dimensional growth, we grow the same stack on 1 ML of MoSe$_2$ grown by co-deposition as described in the previous section. The evolution of the RHEED pattern upon annealing is shown in Fig.~\ref{Fig6}. Up to 600$^{\circ}$C, it exhibits broad rings corresponding to the nucleation of nanometer-sized MoSe$_2$ grains with random orientation at the interface between Mo and the amorphous Se layer. When increasing the temperature, the RHEED pattern progressively evolves towards a streaky pattern corresponding to two-dimensional MoSe$_2$. Hence the MoSe$_2$ grains gradually rotate to lie flat onto the MoSe$_2$ buffer monolayer at 750$^{\circ}$C. Moreover, after the rings disappear, the streaks become thinner which means that the MoSe$_2$ layers are forming a flat and smooth surface. The grazing incidence x-ray diffraction spectra in Fig.~\ref{Fig7} indicate that the MoSe$_2$ film is in tensile strain by +0.2 \% with respect to the bulk lattice parameter.


In both SPE growths, nanometer-sized MoSe$_2$ grains or flakes with random orientation nucleate at the interface between Mo and the amorphous Se layer. When increasing the temperature, Mo atoms are fully consumed to form MoSe$_2$ flakes. Those flakes softly land onto the substrate with a random tilt angle with respect to the surface normal while the excess of Se beneath progressively evaporates. Their edges being highly reactive, when they come into contact with the SiO$_2$ surface they form chemical bonds which prevents the flakes to rotate and lie flat onto the surface. Conversely, when grown on the MoSe$_2$ buffer monolayer, the edges come into contact with a van der Waals surface with no dangling bonds. In this case, the flakes are free to rotate and minimize the system energy by lying parallel to the surface and forming grain boundaries to saturate the dangling bonds at the edges. Using a chemical route, a similar approach was used by G\''ohler \textit{et al.} for the growth of nanocrystalline MoSe2 on epitaxial graphene on SiC \cite{Gohler2018}.

\begin{figure}[hbtp]
\centering
\includegraphics[scale=0.4]{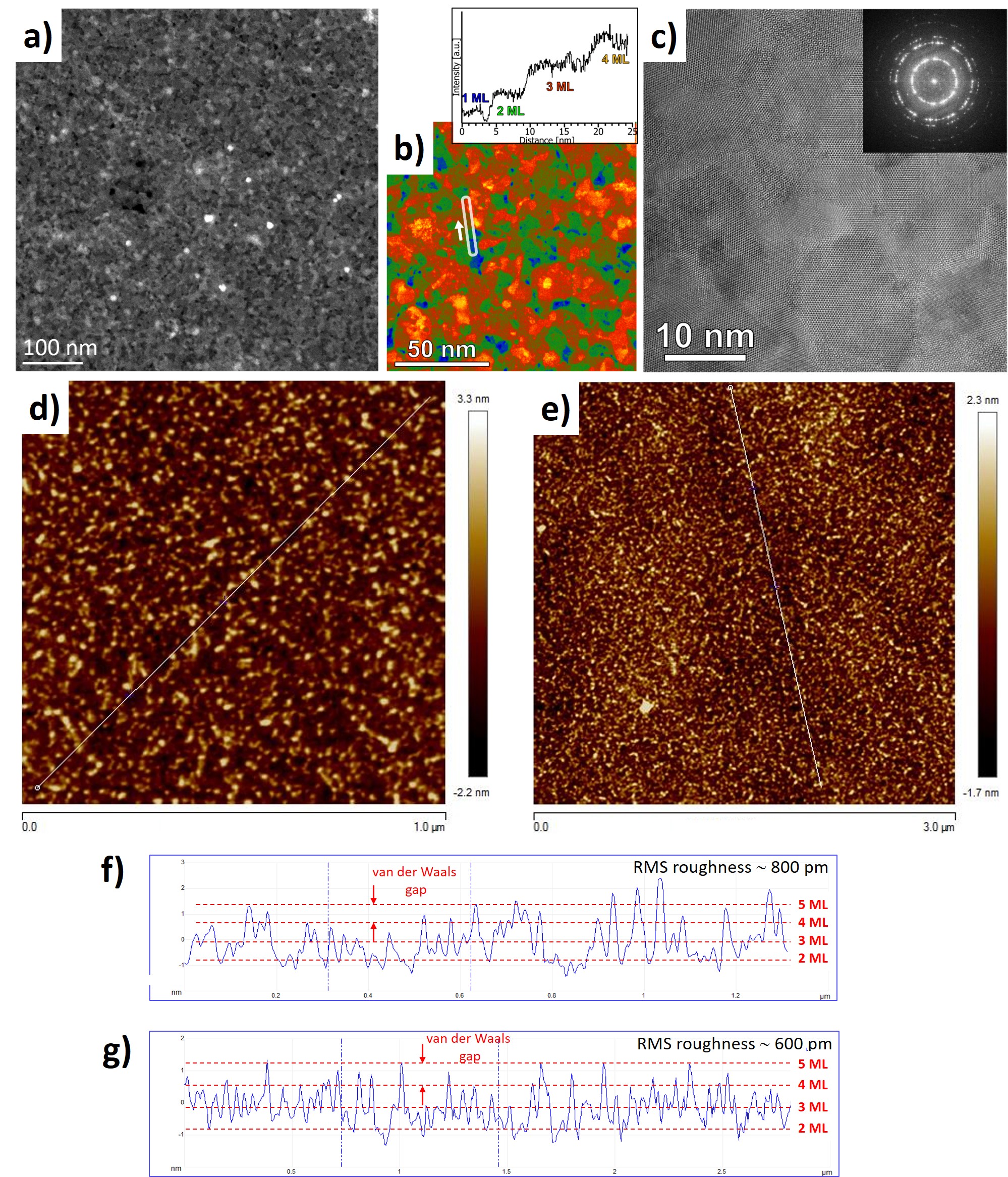}
\caption{\textit{(a)}, STEM image of the 2 ML of MoSe$_2$ layers grown by SPE on top of 1 ML of MoSe$_2$ grown by co-deposition. \textit{(b)} False color STEM image. Blue color corresponds to 1 ML, green color to 2 ML, red color to 3 ML and yellow color to 4 ML. Inset: height profile recorded along the white rectangle indicated. \textit{(c)}, TEM image and corresponding electron diffraction pattern. \textit{(d)}, 1$\times$1 $\mu$m$^{2}$ and \textit{(e)}, 3$\times$3 $\mu$m$^{2}$ AFM images. \textit{(f)} and \textit{(g)}, Height profiles recorded along the white solid lines in \textit{(d)} and \textit{(e)} respectively. The root mean square (RMS) roughness and the expected distance between two MoSe$_2$ layers (the van der Waals gap of 650 pm) are indicated in the height profiles. Due to the presence of the co-deposited MoSe$_2$ monolayer underneath, the heights start at 2 ML.}\label{Fig8}
\end{figure}
\begin{figure}[hbtp]
\centering
\includegraphics[scale=0.5]{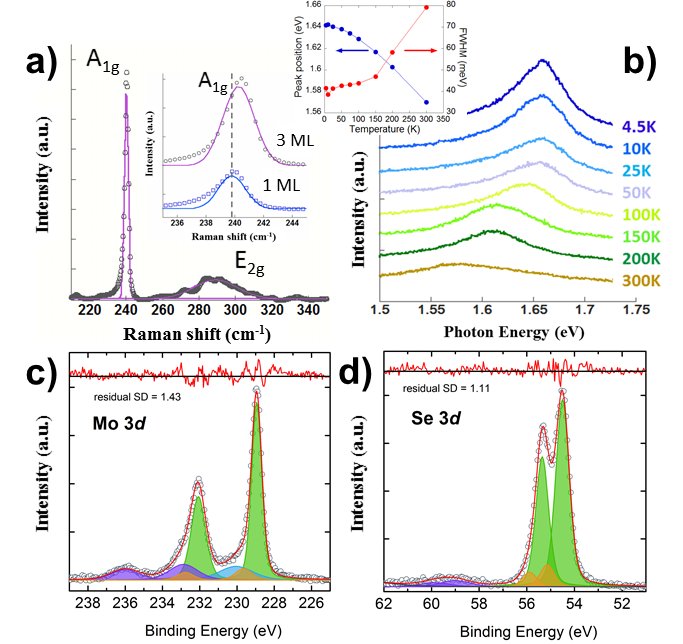}
\caption{\textit{(a)}, Typical Raman spectrum obtained on 2 ML of MoSe$_2$ grown by SPE on top of 1 ML of MoSe$_2$ grown by co-deposition. The total thickness is 3 ML. Inset: comparison with the $A_{1g}$ peak position of 1 ML of MoSe$_2$ grown by co-deposition. The pink and blue solid lines are fits of the $A_{1g}$ peaks. \textit{(b)}, Photoluminescence signal as a function of temperature. Inset: position and FWHM of PL peaks as a function of temperature. XPS Mo 3d \textit{(c)} and Se 3d \textit{(d)} high-resolution core level spectra. }\label{Fig9}
\end{figure}

The monolayer of MoSe$_2$ acts as a van der Waals template and favors the two-dimensional growth of the two topmost MoSe$_2$ layers grown by SPE. We thus call this growth method van der Waals SPE. The STEM images in Fig.~\ref{Fig8}a and ~\ref{Fig8}b show that the total film thickness is actually 3 ML$\pm$1 ML. A height profile is shown in the inset of Fig.~\ref{Fig8}b. In Fig.~\ref{Fig8}c, the TEM image along with its Fourier transform show the polycrystalline character of the film with a grain size comparable to the one of the co-deposited MoSe$_2$ monolayer i.e. 10-50 nm. AFM images in Fig.~\ref{Fig8}d and ~\ref{Fig8}e demonstrate the uniformity of the MoSe$_2$ over micrometers and are consistent with STEM images. The corresponding height profiles are shown in Fig.~\ref{Fig8}f and ~\ref{Fig8}g and the film roughness corresponds approximately to the sum of the substrate roughness ($\approx$200 pm) and 1 ML of MoSe$_2$ (650 pm). A typical Raman spectrum of the SPE-grown MoSe$_2$ layers is shown in Fig.~\ref{Fig9}a. We still observe the $A_{1g}$ and $E_{2g}$ vibration modes of MoSe$_2$ at 241.1 cm$^{-1}$ and 288.9 cm$^{-1}$ respectively. 


The $A_{1g}$ peak position is shifted by +1.3 cm$^{-1}$ for 3 ML of MoSe$_2$. Following Ref. \cite{Tonndorf2013}, from 1 ML to 2 ML (resp. 3 and 4 ML) of MoSe$_2$ the shift of the $A_{1g}$ Raman peak position is +1 cm$^{-1}$ (resp. +2 cm$^{-1}$). Our shift value is intermediate and can be explained by the thickness dispersion shown in Fig.~\ref{Fig8}b by STEM. Indeed, 2 ML represent 27 \% of the total surface while 3 and 4 ML represent 73 \%. By considering that the $A_{1g}$ Raman peak intensity of 2 ML is 2-3 times larger than the ones of 3 and 4 ML \cite{Tonndorf2013}, we end with a peak shift of +1.8 cm$^{-1}$. This value (which is an upper bound since we did not take into account the 1 ML coverage corresponding to the template co-deposited layer) is in reasonable agreement with our experimental value. The $A_{1g}$ peak width is of the order of 2.5 cm$^{-1}$ which is larger than the one of 3 ML exfoliated flakes ($\approx$1.5 cm$^{-1}$) \cite{Tonndorf2013}. This enlargement is probably due to the thickness dispersion of our film. We also performed Raman maps over 20$\times$20 $\mu$m$^{2}$ areas with a 2 $\mu$m step, the spot size being of the order of 1 $\mu$m. All the recorded peaks fall at the same position and exhibit the same intensity within 5 \% showing the good uniformity of the MoSe$_2$ film over large areas. The temperature dependence of the photoluminescence is shown in Fig.~\ref{Fig9}b. The corresponding peak position and FWHM are displayed in the inset. The PL spectra are very similar to those of the co-deposited MoSe$_2$ monolayer shown in Fig.~\ref{Fig3}b except that the peak position is systematically shifted by 20 meV to lower energy and the PL intensity (in the same excitation conditions) is one order of magnitude less. The same observations were reported for MoSe$_2$ exfoliated flakes \cite{Tonndorf2013} and correspond precisely to the transition from a direct bandgap in 1 ML MoSe$_2$ to an indirect bandgap in MoSe$_2$ multilayers. It demonstrates that MBE grown MoSe$_2$ layers exhibit the same optical properties as exfoliated ones. At 4.5 K, the photoluminescence peak width of $\approx$100 meV is larger than the one of 1 ML MoSe$_2$ which reflects the thickness dispersion like in Raman spectroscopy. Finally, the temperature dependence of the FWHM for 1 and 3 ML superimpose suggesting again the existence of a second PL peak at lower energy corresponding to the trion emission in the 3 ML-thick film below 150 K. The Mo 3d and Se 3d high-resolution XPS spectra of 3 ML MoSe$_2$ are displayed in Fig.~\ref{Fig9}c and ~\ref{Fig9}d respectively. As discussed in the previous section, the presence of oxidized defects in the layers adds secondary doublets in the Mo 3d and Se 3d spectra. Their relative contributions are given in Table~\ref{Table1}. Stoichiometric MoSe$_2$ in green is measured at 228.9 eV for Mo 3d$_{5/2}$ and 54.5 for Se 3d$_{5/2}$ in good agreement with the spectra presented in Fig.~\ref{Fig3}c and ~\ref{Fig3}d respectively i.e. 232.9 eV and 59.0 eV respectively. The oxidized defects (MoSe$_x$O$_y$ in orange, MoO$_x$ and SeO$_x$ in purple) are at the same energy positions as for the co-deposited MoSe$_2$ monolayer. The total amount of oxidized defects is almost the same as the one in the co-deposited MoSe$_2$ monolayer. We find 3 \% (resp. 3 \%) for 3 ML (resp. 1 ML) for Mo 3d spectra and 10 \% (resp. 13 \%) for 3 ML (resp. 1 ML) for Se 3d spectra. However, we note a significant reduction of the oxide doublets for Mo (19 \% instead of 25 \%) and Se (6 \% instead of 7 \%). This observation can be easily explained by the fact that buried MoSe$_2$ layers are protected against deep oxidation. However, the fact that the total amount of defects remain constant means that the density of point defects is probably higher by SPE. It could be due to the rapid evaporation of Se during the annealing process and to the resulting formation of Se vacancies.

\begin{figure}[hbtp]
\centering
\includegraphics[scale=0.45]{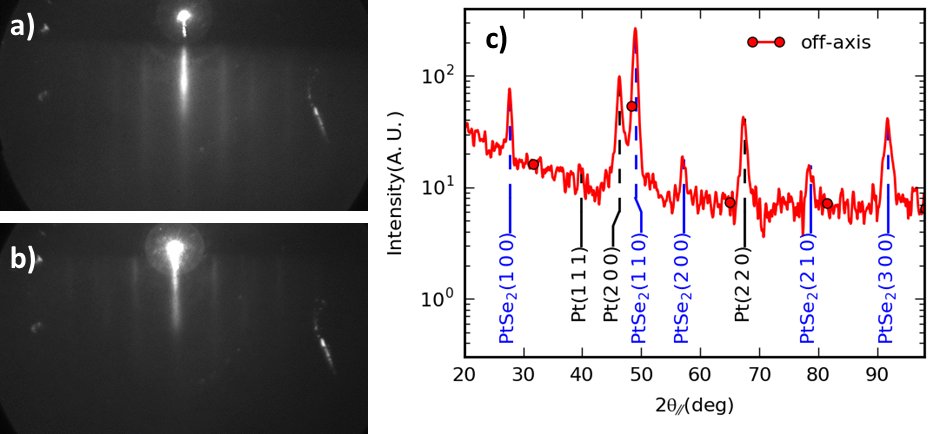}
\caption{\textit{(a)} and \textit{(b)} RHEED patterns of PtSe$_2$ and SnSe$_2$ grown by vdW epitaxy on 1 ML of MoSe$_2$ grown by co-deposition on SiO$_2$/Si. For PtSe$_2$, 10 nm of amorphous Se was first deposited with 2 nm of pure platinum on top. Well-defined streaks in the RHEED pattern are obtained after annealing at  600$^{\circ}$C. Due to the excess of platinum, some rings are visible corresponding to pure platinum particles on top of PtSe$_2$. For SnSe$_2$, 10 nm of amorphous Se was first deposited with 1 nm of pure tin on top. Well-defined streaks in the RHEED pattern are also obtained after annealing at 300$^{\circ}$C. \textit{(c)} Grazing incidence x-ray diffraction spectrum of PtSe$_2$ along a radial reciprocal space direction (11.90$^{\circ}$ off the Si(hh0) crystal axis). Some platinum peaks are also visible due to the excess of metal with respect to selenium. Vertical dotted lines indicate the different peak positions.}\label{Fig10}
\end{figure}

In conclusion, this vdW-SPE growth technique allowed us to grow good quality MoSe$_2$ mutilayers. We could also grow PtSe$_2$ and SnSe$_2$ multilayers (as shown in Fig.~\ref{Fig10}) using the same technique.

\section{Mn-doped MoSe$_2$ grown by vdW-SPE}

\begin{figure}[hbtp]
\centering
\includegraphics[scale=0.35]{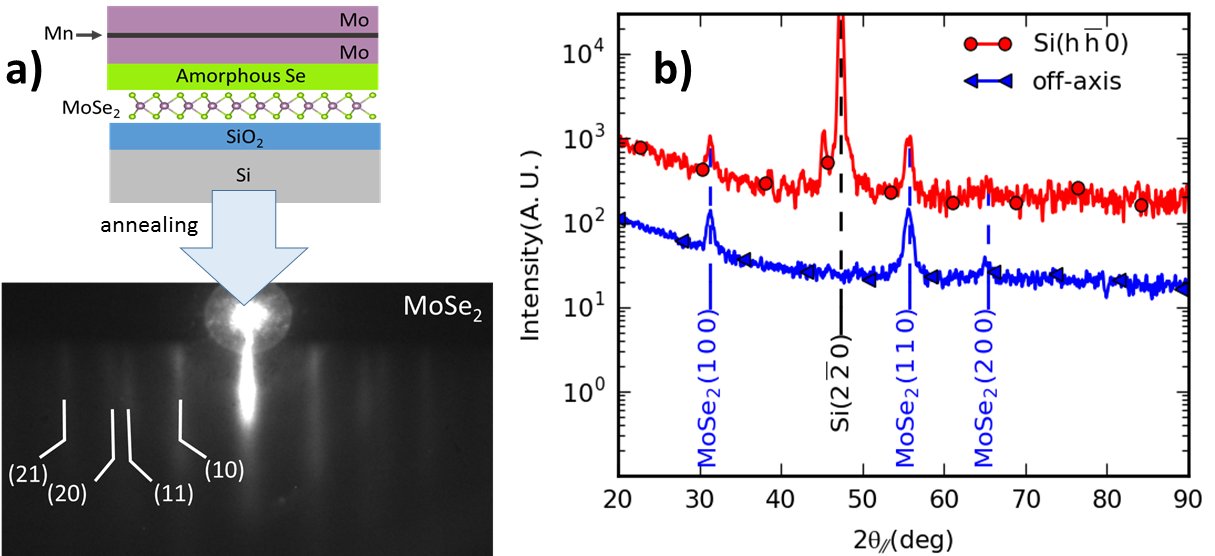}
\caption{\textit{(a)} Illustration of the layer stack to dope MoSe$_2$ with Mn by vdW-SPE and RHEED pattern obtained after annealing. Again the initial RHEED pattern is diffuse showing the amorphous character of the stack. \textit{(b)} X-ray diffraction spectra of MoSe$_2$ doped with 10 \% of Mn by vdW-SPE along two different radial reciprocal space directions showing the isotropic character of the film. The vertical dotted lines indicate the different peak positions.}\label{Fig11}
\end{figure}

\begin{figure}[hbtp]
\centering
\includegraphics[scale=0.4]{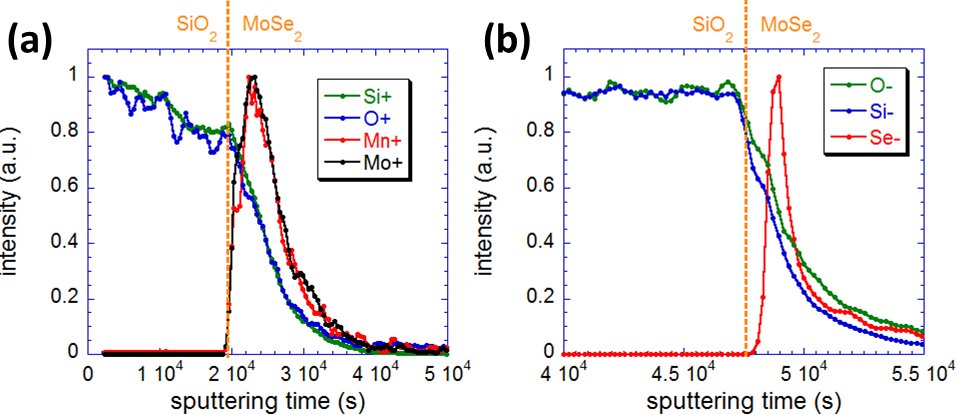}
\caption{\textit{(a)}, TOF-SIMS profiles of Si$^{+}$, O$^{+}$, Mo$^{+}$ and Mn$^{+}$ ions. The sample is sputtered from the backside in order to study the diffusion of atomic species (Mo, Mn and Se) in the SiO$_2$ substrate. \textit{(b)}, TOF-SIMS profiles of O$^{-}$, Si$^{-}$ and Se$^{-}$ ions. The orange vertical dotted lines indicate the interface between SiO$_2$ and MoSe$_2$:Mn.}\label{Fig12}
\end{figure}

\begin{figure}[hbtp]
\centering
\includegraphics[scale=0.35]{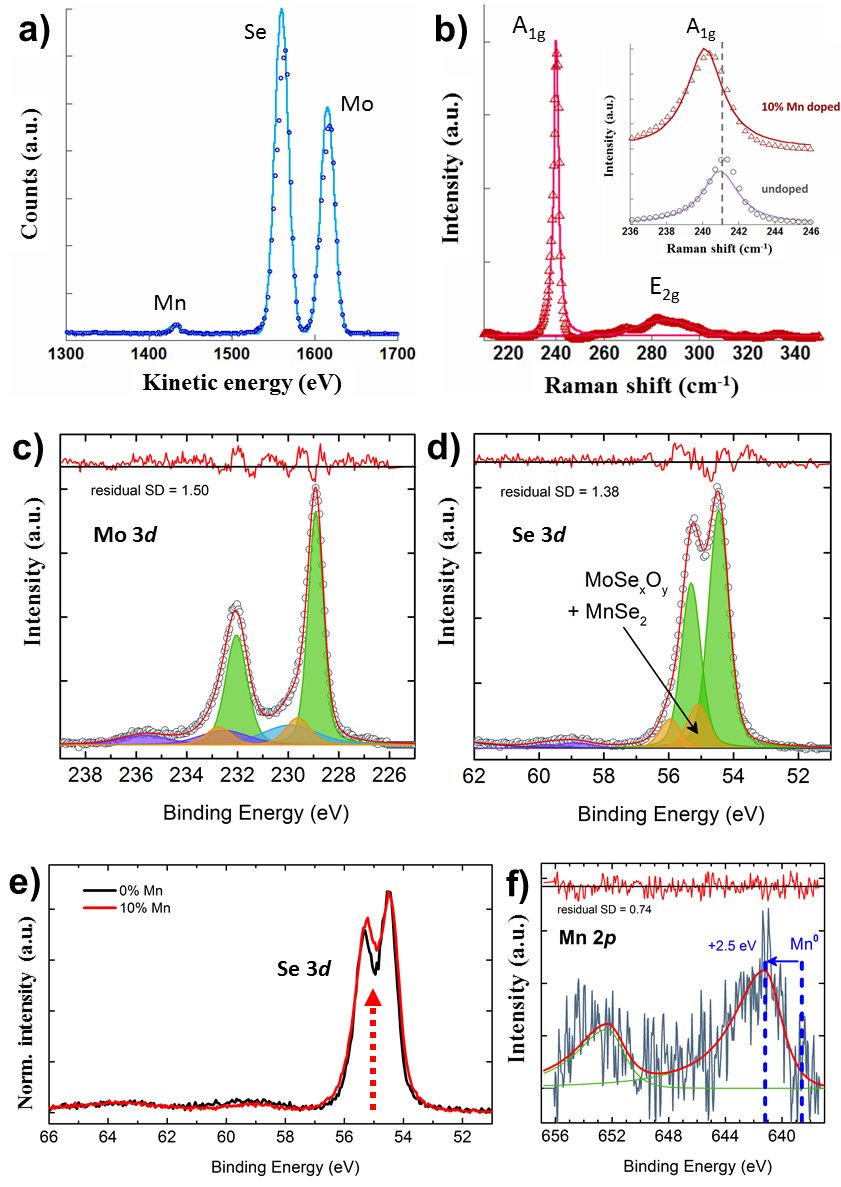}
\caption{\textit{(a)}, RBS spectrum obtained on the Mo$_{0.91}$Mn$_{0.09}$Se$_2$ film grown by vdW-SPE. The blue open dots are experimental data and the blue solid line is the fitting curve. \textit{(b)}, Typical Raman spectrum obtained on 2 ML of MoSe$_2$ grown by SPE and doped with 10 \% of Mn on top of 1 ML of MoSe$_2$ grown by co-deposition. Inset: comparison with the $A_{1g}$ peak position of 3 ML undoped MoSe$_2$. The red and purple solid lines are fits of the $A_{1g}$ peaks. \textit{(c)}, and \textit{(d)}, XPS spectra of Mo 3d and Se 3d core levels respectively. \textit{(e)}, Comparison of Se 3d core level XPS spectra with and without Mn. The red dotted arrow points at the emergence of a new peak upon Mn doping at a binding energy close to 55 eV which corresponds to the Mn-Se chemical bond. \textit{(f)}, XPS spectrum of Mn 2p core levels.}\label{Fig13}
\end{figure}

At last, as illustrated in Fig.~\ref{Fig11}a, we use this vdW-SPE technique to dope the two topmost MoSe$_2$ layers with manganese. In MoS$_2$, a Mn atomic concentration of 10 \% could lead to two-dimensional ferromagnetism \cite{Cheng2013}. Some experimental attempts to dope MoS$_2$ with Mn have also been reported in nanostructures \cite{Wang2016} or monolayers \cite{Zhang2015}. In MoSe$_2$, when a Mn atom substitutes a Mo atom, it acts as a donor and a magnetic impurity of moment 1.27 $\mu_{B}$ \cite{Mishra2013}. Ferromagnetism is then predicted in Mo$_{1-x}$Mn$_x$Se$_2$ for $x>$0.05 \cite{Mishra2013}. For this reason, we insert a 0.06 nm-thick Mn layer in between two 0.3 nm-thick Mo layers to grow Mo$_{0.91}$Mn$_{0.09}$Se$_2$ (here the subscripts are for volume fractions). If we consider the molar volumes of Mo and Mn, we find the atomic composition Mo$_{0.887}$Mn$_{0.113}$Se$_2$. The annealing process leads to the same RHEED pattern as the one of the undoped film as shown in Fig.~\ref{Fig11}a. Using this vdW-SPE technique, manganese is not in direct contact with SiO$_2$. It was shown that Mn atoms strongly react with the oxide layer which is detrimental to its incorporation into MoS$_2$ \cite{Zhang2015}. STEM and TEM observations give the same layer morphology and grain size. There is no evidence for clustering. We performed grazing-incidence x-ray diffraction (Fig.~\ref{Fig11}b) and found no secondary phase. The deduced lattice parameter of MoSe$_2$ corresponds to the bulk one with a +0.3 \% tensile strain very close to the one of the undoped film (+0.2 \%). Considering that this tensile strain amplitude is close to the resolution of our setup, we conclude that doping MoSe$_2$ with up to 10 \% of Mn by vdW-SPE does not change drastically the lattice parameter. We also find a lower bound for the grain size of 16 nm. We perform TOF-SIMS. A first set of experiments (not shown) was carried out by sputtering the sample from the frontside. We measured Mo$^{+}$, Mn$^{+}$ and Se$^{-}$ species which are the elements of interest. The decreasing signal tails observed towards the bulk could suggest a possible diffusion in the SiO$_2$ underlayer knowing that ion mixing can produce the same effect. To circumvent this difficulty, a backside sample preparation was performed. It is based on the silicon substrate removal combining mechanical polishing and chemical etching. The backside analyses of the resulting SiO$_2$/MoSe$_2$:Mn structure do not reveal any trace of Mo, Mn (Fig.~\ref{Fig12}a) and Se (Fig.~\ref{Fig12}b) in the oxide layer. The results also confirm that Mn atoms are well incorporated into MoSe$_2$: the Mo and Mn profiles perfectly superimpose. We then confirm the stoichiometry using RBS, the spectra are shown in Fig.~\ref{Fig13}a. Since the substrate contains only light atoms (Si and O), we can easily distinguish the Mn, Se and Mo peaks in Fig.~\ref{Fig13}a. By taking into account the diffusion cross section of each atomic species and assuming that Mn atoms are substituting Mo atoms in the two topmost layers, we fit the three peaks as shown in Fig.~\ref{Fig13}a. We obtain the following atomic composition Mo$_{0.88\pm 0.03}$Mn$_{0.12\pm 0.03}$Se$_2$ which is in good agreement with the composition target within error bars.


The Raman spectrum of the Mn-doped MoSe$_2$ film is displayed in Fig.~\ref{Fig13}b. The $A_{1g}$ (resp. E$_{2g}$) peak position is 240.1 cm$^{-1}$ (resp. 285.6 cm$^{-1}$). Compared to the undoped film in the inset of Fig.~\ref{Fig13}b, the A$_{1g}$ peak position is shifted by -1 cm$^{-1}$ but its width is the same (2.4 cm$^{-1}$). The shift to lower wave numbers has already been observed in n-doped MoS$_2$ \cite{Pham2016} which may be compatible with the donor character of Mn substituting Mo in MoSe$_2$. This doping effect would require additional electrical measurements that are out of the scope of the present work. However, in Fig.~\ref{Fig13}c and ~\ref{Fig13}d, the Mo 3d$_{5/2}$ and Se 3d$_{5/2}$ peaks of the stoichiometric MoSe$_2$ phase in green are at the same energy positions as the ones of the undoped sample \textit{i.e.} 228.9 eV and 54.5 eV respectively. Hence we do not detect a significant binding energy shift upon Mn doping meaning that Mn incorporation does not induce an electronically active n-type doping in MoSe$_2$. The peak shift observed in the Raman spectrum has thus another physical meaning. It may be due to strain effects induced by the presence of Mn atoms within individual MoSe$_2$ layers in substitutional position \cite{Azcatl2016}. However, within our setup resolution, grazing incidence x-ray diffraction in Fig.~\ref{Fig11}b shows no variation of the lattice parameter. It may also be due to the possible intercalation of Mn atoms between MoSe$_2$ sheets. At this point, further work is needed to give a reliable conclusion. Concerning the additional doublets observed in XPS spectra (MoSe$_x$O$_y$ in orange and MoO$_x$, SeO$_x$ in purple), the two main evolutions due to Mn doping are: \textit{(i)} the lower fraction of oxide components: 16 \% instead of 19 \% for Mo 3d and 3 \% instead of 6 \% for Se 3d and \textit{(ii)} a significant increase of the MoSe$_x$O$_y$ component: 5 \% instead of 3 \% for Mo 3d and 16 \% instead of 10 \% for Se 3d. If we assume that Mn atoms are present at the edges of MoSe$_2$ grains, it reduces the fraction of oxidized Mo and Se atoms. It also means that the Mn-O bonding cannot be neglected in the Mn 2p peaks as discussed below. Concerning the fraction of MoSe$_x$O$_y$, we notice a sharp increase of its contibution on Se 3d peaks. Following Ref. \cite{Wang2006}, the Se 3d binding energy corresponding to the Mn-Se bond in cubic MnSe$_2$ is 55.15 eV which corresponds to the peak position of Se 3d in MoSe$_x$O$_y$. As shown in Fig.~\ref{Fig13}e, the sharp increase of this peak intensity (+6 \%) is thus due to the presence of Mn-Se bonds and we can conclude that a fraction of Mn atoms substitute Mo atoms in the MoSe$_2$ layers. Since the formation of Mn-Se bonds represent 6 \% of the total amount of Se atoms, we can roughly estimate that 3 \% of Mo atoms are substituted by Mn atoms. The rest of the Mn atoms might be in intercalation between MoSe$_2$ sheets or more probably at the edges of MoSe$_2$ grains. In Fig.~\ref{Fig13}f, we show the Mn 2p core level peaks. The very low intensity due to our experimental conditions makes any interpretation difficult. Nevertheless, we can identify easily the two spin-orbit components at 641.2 eV and 652.7 eV, but we cannot state that the spectrum corresponds to a unique doublet or to a sum of different bonding-states. However, this doublet presents a 2.5 eV shift to higher binding energies compared to metallic Mn, indicated by Mn$^{0}$ in the figure. The 2p$_{3/2}$ peak position at 641.2 eV may correspond to the Mn-Se bonding as expected (640.5-641.0 eV in Ref. \cite{Mirabella2004} and 640.95 eV in Ref. \cite{Wang2006}), but it could also be a Mn-O signature which is very close in terms of binding energy \cite{Choi1996}. From a quantitative analysis of XPS spectra taking into account all the Mo components (including the ones from the undoped MoSe$_2$ template layer), we find the following atomic composition: Mo$_{0.89\pm 0.03}$Mn$_{0.11\pm 0.03}$Se$_2$ which is again very close to the composition target. We have thus successfully incorporated $\approx$10 \% of Mn atoms in MoSe$_2$. This vdW-SPE technique can be used for other dopants such as electrical dopants (Nb, Re...) or other magnetic dopants (V, Cr...). It represents a very promising route to dope TMDC layers on large scales.

\section{Conclusion}

To summarize, we have successfully grown large-area single and multi-layers of MoSe$_2$ on SiO$_2$/Si using two different techniques. By co-depositing Mo and Se atoms by molecular beam epitaxy in the van der Waals regime on SiO$_2$, we could obtain a single MoSe$_2$ layer continuous over $\approx$1 cm$^2$. Despite the small size of individual MoSe$_2$ grains (10-50 nm) and their random in-plane orientation, the MoSe$_2$ layer exhibits the same electronic properties as exfoliated micron-sized MoSe$_2$ flakes. However, using the same growth technique to grow few layers, vertical growth starts to occur at the junction between grains exhibiting a large tilt angle. To circumvent this issue, we developed an original growth method based on the van der Waals solid phase epitaxy (vdW-SPE). It consists in first depositing an amorphous Se/Mo bilayer on top of a co-deposited MoSe$_2$ monolayer. After annealing this stack, we obtain two-dimensional MoSe$_2$ multilayers over 1 cm$^2$. Again these multilayers grown by vdW-SPE exhibit the same electronic properties as the exfoliated multilayered flakes. Moreover, by inserting a thin layer of manganese in the middle of the Mo film, we could demonstrate the incorporation of up to 10 \% of Mn in MoSe$_2$ three-layers. This technique opens a new route to dope TMDC and can be easily extended to other electrical (Nb, Re) or magnetic (V, Cr) dopants. Mn-doped MoSe$_2$ is expected to exhibit ferromagnetic properties and vdW-SPE should allow to explore this new field of two-dimensional magnetism.

\section{Aknowledgements}

The authors would like to thank Vincent Mareau, Laurent Gonon and Joao Paulo Cosas Fernandes for their assistance with Raman spectroscopy and Xavier Marie for helpful discussions on photoluminescence measurements. This work is supported by the CEA-project 2D FACTORY, Agence Nationale de la Recherche within the ANR MoS2ValleyControl (ANR-14-CE26-0017) and ANR MAGICVALLEY projects. The French state funds LANEF (ANR-10-LABX-51-01) is acknowledged for its support with mutualized infrastructure. This work, done on the NanoCharacterisation PlatForm (PFNC), was supported by the Recherches Technologiques de Base Program of the French Ministry of Research. M.T. Dau acknowledges the support from the CEA-Enhanced Eurotalents program.

\appendix

\section*{Appendix: Characterization Methods}

A double aberration-corrected FEI Titan Ultimate is used to conduct the transmission electron microscopy (TEM) and scanning TEM (STEM) experiments. The microscope is operated at 80 kV for both TEM and STEM image acquisition. For TEM imaging, the monochromator is excited to reduce the energy spread of the electron beam to $\approx$0.15 eV. STEM images are acquired using a convergence angle of 21 mrad and camera length of 115 mm. Before transferring the film onto a copper grid for transmission electron microscopy, the MoSe$_2$ layer is covered with graphene. The stack is then transferred from the SiO$_2$(300 nm)/Si substrate using a 2 M KOH solution. The transfer is conducted using a protective polymer (polystyrene), which was deposited via spin coating at 3000 rpm. After detaching the specimen from the growth substrate, the specimen is rinsed several times using deionized water, and transferred to a C-Flat TEM grid with 2 $\mu$m holes. The C-Flat grid with the MoSe$_2$ layer are then submerged in toluene to remove the polystyrene protective film. \\
Atomic force microscopy (AFM) measurements were performed in the peak force mode with a Bruker Dimension-Icon setup operated in ambient conditions, with a scanasyst-air cantilever (nominal spring constant 0.4 Nm$^{-1}$). The force set-point was automatically adjusted by the controller which was operated in ScanAsyst mode (peak force amplitude: 40 nm, $z$-piezo oscillation frequency: 2.0 kHz).\\
Confocal Raman spectroscopy is conducted in backscattering mode using a LabRam HR (laser He-Ne at 632.8 nm - 17 mW), with a $\times$100 long working distance objective. The spot size is approximately 1 $\mu$m, the acquisition time is 20 secondes per spectrum and we systematically accumulate 3 spectra to improve the signal-to-noise ratio. \\
For photoluminescence measurements, we used a He-Ne laser ($\lambda$=632.8 nm) as the source of excitation with a power of 2 mW. The diameter of the laser spot on the sample is 1.45 $\mu$m and the light is collected through a $\times$50 objective of numerical aperture 0.55. The acquisition time is 40 secondes per spectrum.\\
XPS measurements were performed under ultrahigh vacuum with an X-ray microprobe (beam size 100 $\mu$m) photoelectron spectrometer (PHI 5000 VersaProbe-II) fitted with a monochromatic Al K$_\alpha$ source (h$\nu$ = 1486.7 eV). The samples were transferred in air prior to the measurements. The photoelectron take-off angle was 45$^{\circ}$ and the overall energy resolution 0.6 eV for high-resolution core level spectra. The carbon 1s peak, commonly used for energy referencing, overlaps with a selenium Auger line and becomes difficult to identify when the sample is poorly covered by adventitious carbon \cite{Weightman1975}. Therefore, the energy scale was calibrated by shifting the silicon 2p peak, coming from the SiO$_2$(300 nm)/Si substrate, to 103.7 eV \cite{XPS_SiO2}. The decomposition of the core-level line shape into individual bonding-state components was done with the CasaXPS software, using 30 \%-Lorentzian/Gaussian sum line-shapes and Shirley background subtraction. The spin-orbit splitting was set to 3.1 eV for Mo 3d, 0.9 eV for Se 3d. The relative concentration of each element was calculated using the corrected relative sensitivity factor given by the instrument, considering the geometric and experimental specificities. For Mn 2p core levels, the spin-orbit splitting was set to 11.5 eV. We considered a Lorentzian asymmetric line shape according to Ref. \cite{Biesinger2011} and a linear background due to a poor peak intensity caused by a low photoionization cross section \cite{Lindau1985} and a small amount of atoms combined with a strong inelastic background from the nearby Se Auger lines. \\
X-ray diffraction measurements were performed with a SmartLab Rigaku diffractometer equipped
with a copper rotating anode beam tube ($\lambda_{K\alpha}$=1.5405 \AA) and operated at 45 kV and 200 mA.\\
The in-depth elemental distribution was measured by Time-of-Flight Secondary Ions Mass Spectrometry (TOF-SIMS) on the ION TOF V dual beam instrument. The positive depth profiles were acquired using a 1 keV O$^{2+}$ sputter beam (230 nA, raster 150$\times$150 $\mu$m$^{2}$) and a pulsed 25 keV Bi$^{+}$ analysis beam rastered over a 40$\times$40 $\mu$m$^{2}$ area. Both beams were incident at approximately 45$^{\circ}$. In addition, the negative depth profiles were obtained with the same 25 keV Bi$^{+}$ for analysis (raster 60$\times$60 $\mu$m$^{2}$) and using a 2 keV Cs$^{+}$ sputter beam (70 nA, raster 200$\times$200 $\mu$m$^{2}$). An electron flood gun was used for charge compensation needed for the investigation of the insulating oxide layer. The secondary ions ejected from the sample surface are accelerated toward a detector and their mass-to-charge ratio is determined by a time of flight measurement.\\
The Rutherford backscattering spectrometry (RBS) technique is based on the elastic scattering of 2 MeV $^{4}$He$^{+}$ ions by the atomic nuclei of the sample. The pressure in the analysis chamber is 2$\times$10$^{-6}$ Torr and the sample current is 4 nA, the incident ion beam is few mm$^{2}$ large. The angle between the incident beam and the detector is 160$^{\circ}$ and the angle of incidence is set to 75$^{\circ}$ with respect to the sample normal in order to limit channeling effects and enhance the apparent film thickness probed by the beam.

\cleardoublepage

\cleardoublepage

\cleardoublepage

\cleardoublepage

\cleardoublepage

\cleardoublepage

\cleardoublepage

\cleardoublepage

\cleardoublepage

\cleardoublepage

\cleardoublepage

\cleardoublepage

\cleardoublepage

\end{document}